\documentclass[sigconf]{acmart}
\AtBeginDocument{%
  }

\setcopyright{acmlicensed}
\copyrightyear{2018}
\acmYear{2018}
\acmDOI{XXXXXXX.XXXXXXX}
\acmConference[Conference acronym 'XX]{Make sure to enter the correct
  conference title from your rights confirmation email}{June 03--05,
  2018}{Woodstock, NY}
\acmISBN{978-1-4503-XXXX-X/2018/06}

\usepackage{multirow}

\renewcommand\footnotetextcopyrightpermission[1]{} 
\begin{document}

\title{NarraScore: Bridging Visual Narrative and Musical Dynamics via Hierarchical Affective Control}

% \author{Member A}
% % \authornote{Both authors contributed equally to this research.}
% \email{somebody@mails.tsinghua.edu.cn}
% \affiliation{%

\author{Yufan Wen}
\affiliation{%
  \institution{Tsinghua University}
  \city{Shenzhen}
  \country{China}}
\email{wenyf24@mails.tsinghua.edu.cn}

\author{Zhaocheng Liu}
\authornote{Corresponding author.}
\affiliation{%
  \institution{ByteDance}
  \city{Beijing}
  \country{China}
}
\email{lio.h.zen@gmail.com}

\author{YeGuo Hua}
\affiliation{%
  \institution{ByteDance}
  \city{Beijing}
  \country{China}
}
\email{huayeguo@bytedance.com}

\author{Ziyi Guo}
\affiliation{%
  \institution{ByteDance}
  \city{Shenzhen}
  \country{China}
}
\email{ziyi.94@bytedance.com}

\author{Lihua Zhang}
\affiliation{%
  \institution{ByteDance}
  \city{Beijing}
  \country{China}
}
\email{lizhiyu.0@bytedance.com}

\author{Chun Yuan}
\authornotemark[1]
\affiliation{%
  \institution{Tsinghua University}
  \city{Shenzhen}
  \country{China}}
\email{yuanc@sz.tsinghua.edu.cn}

\author{Jian Wu}
\affiliation{%
  \institution{ByteDance}
  \city{Beijing}
  \country{China}
}
\email{wujian@bytedance.com}

\begin{abstract}
  Synthesizing coherent soundtracks for long-form videos remains a formidable challenge, currently stalled by three critical impediments: computational scalability, temporal coherence, and, most critically, a pervasive semantic blindness to evolving narrative logic. To bridge these gaps, we propose NarraScore, a hierarchical framework predicated on the core insight that emotion serves as a high-density compression of narrative logic. Uniquely, we repurpose frozen Vision-Language Models (VLMs) as continuous affective sensors, distilling high-dimensional visual streams into dense, narrative-aware Valence-Arousal trajectories. Mechanistically, NarraScore employs a Dual-Branch Injection strategy to reconcile global structure with local dynamism: a \textit{Global Semantic Anchor} ensures stylistic stability, while a surgical \textit{Token-Level Affective Adapter} modulates local tension via direct element-wise residual injection. This minimalist design bypasses the bottlenecks of dense attention and architectural cloning, effectively mitigating the overfitting risks associated with data scarcity. Experiments demonstrate that NarraScore achieves state-of-the-art consistency and narrative alignment with negligible computational overhead, establishing a fully autonomous paradigm for long-video soundtrack generation.
\end{abstract}

\begin{CCSXML}
<ccs2012>
   <concept>
       <concept_id>10010405.10010469.10010475</concept_id>
       <concept_desc>Applied computing~Sound and music computing</concept_desc>
       <concept_significance>500</concept_significance>
       </concept>
 </ccs2012>
\end{CCSXML}

\ccsdesc[500]{Applied computing~Sound and music computing}

\keywords{Video-to-Music Generation, Long-form Video Soundtrack Generation, Affective Narrative Alignment, Controllable Music Generation}


\maketitle

\section{Introduction}

Background music serves as the emotional pulse of multimedia content, functioning as a narrative engine that actively shapes viewer immersion rather than merely accompanying visuals~\cite{dasovich2022exploring, ma2022research, millet2021soundtrack}. A professional-grade soundtrack is defined by its organic synchronization with visual progression, mirroring the underlying temporal
\begin{figure}[h]
    \centering
    \includegraphics[width=\linewidth]{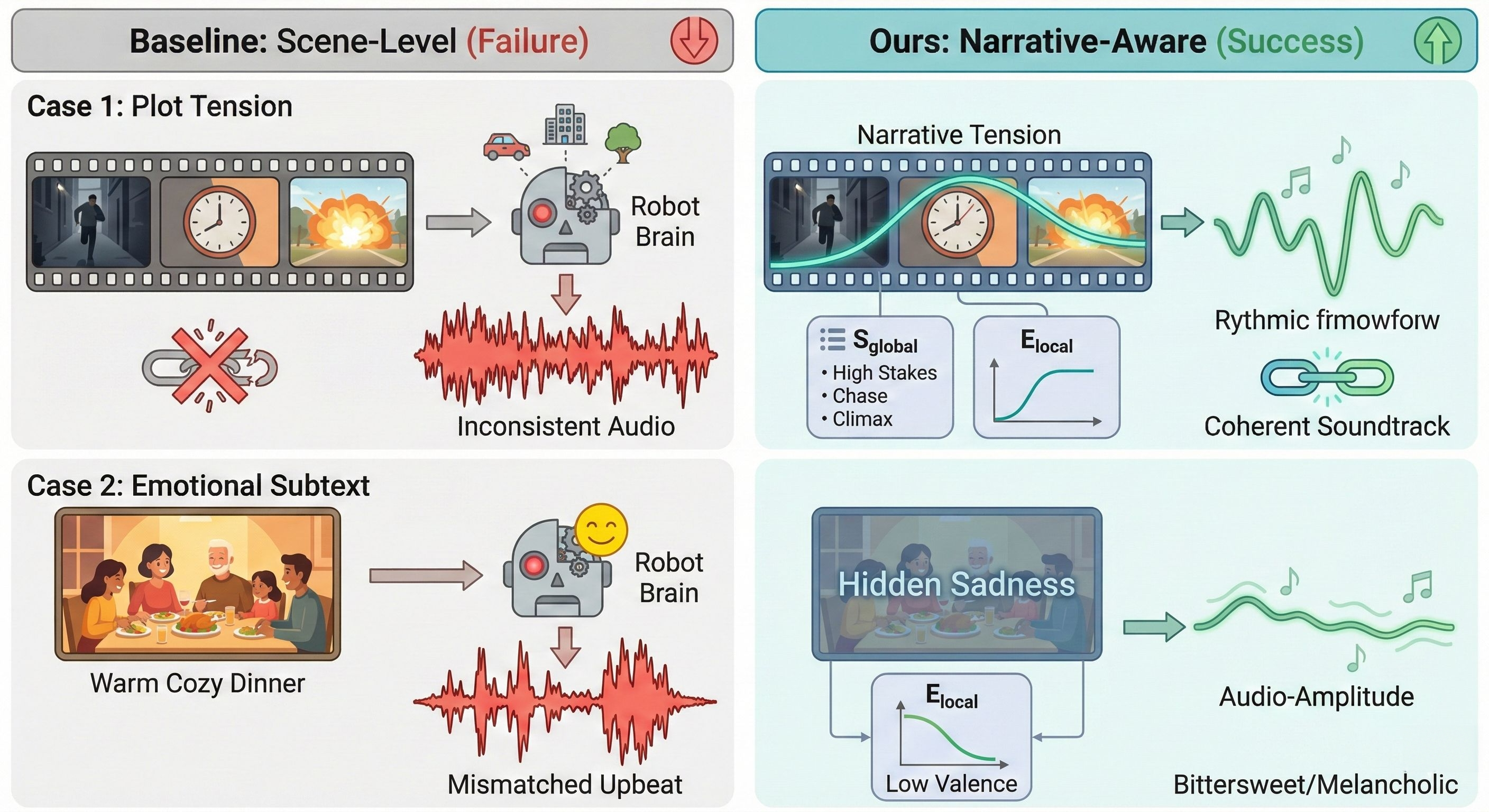}
    \caption{Narrative-aware video background music generation. Unlike baselines that rely on surface visuals and fail to capture narrative tension or hidden subtext, our approach leverages global plot $S_{global}$ and local emotion $E_{local}$ to generate soundtracks that are temporally coherent and narratively resonant.}
    \label{fig:viz}
\end{figure}
dynamics and evolving emotional arcs. With the rapid advancement of generative models\cite{openai2024sora, deepmind2025veo3, wiedemer2025video, liu2025improving} enabling the transition from short clips to long-form video creation, the core challenge has fundamentally shifted, transcending the synthesis of high-fidelity static loops to necessitate the orchestration of complex, evolving soundtracks that maintain stylistic unity while dynamically responding to the shifting pacing and intensity of the visual narrative.

Recent paradigms, while achieving impressive fidelity on short clips via strict frame-wise guidance~\cite{di2021video, gan2020foley, kang2024video2music, su2024v2meow, zuo2025gvmgen, lin2025vmas}, encounter systemic bottlenecks when extrapolated to long-form narratives. Physically, maintaining dense frame-level attention across minute-long sequences incurs prohibitive quadratic memory costs and suffers from attention dilution, where critical narrative cues are drowned out by visual redundancy. Structurally, the absence of global semantic anchors in standard autoregressive models~\cite{agostinelli2023musiclm, huang2018music, copet2023simple} leads to severe style drift, causing the musical identity to fragment over time. Most critically, at a cognitive level, these methods rely on surface-level visual representations~\cite{feichtenhofer2019slowfast, tong2022videomae, wang2023videomae, rouditchenko2020avlnet}, resulting in a semantic blindness—an inability to capture the deep narrative logic, such as rising tension or resolution, which is essential for cinematic storytelling~\cite{ji2025comprehensive}.

To bridge this semantic gap, we propose \textbf{NarraScore}, a hierarchical framework predicated on the core insight that emotion serves as a high-density compression of deep narrative logic. Moving beyond paradigms that depend on unstable external classifiers or expensive manual annotation, we introduce a Lightweight Latent Affective Decoder designed to probe the rich semantic priors encapsulated within frozen Vision-Language Models (VLMs)\cite{zhang2023video, cheng2024videollama, zhang2025videollama}. By explicitly disentangling deep narrative cues from surface-level visual redundancy, this module projects high-dimensional video streams into a compact, autonomous affective manifold. Consequently, NarraScore empowers the system to endogenously deduce evolving emotional arcs directly from raw pixels, establishing a new paradigm of autonomous narrative alignment.

Translating this insight into a generative architecture, NarraScore employs a Dual-Branch Injection strategy to reconcile global coherence with local dynamism. At the macro level, a Global Semantic Anchor conditions the model on the overarching genre and atmosphere, ensuring stylistic stability. At the micro level, we introduce a Token-Level Affective Adapter to modulate finer narrative tension~\cite{zhang2023adding, lan2024musicongen}. Diverging from prevalent paradigms that rely on heavy architectural cloning, we adopt a minimalist projection strategy: distilled continuous affective cues (Valence \& Arousal) are injected as a lightweight additive bias directly into the decoder’s hidden states. This mechanism modulates the semantic manifold via direct additive bias, enabling precise token-level alignment while introducing negligible parameter overhead and preserving the generative priors of the frozen backbone.

In summary, this work makes three key contributions to the field of intelligent content generation. First, we establish a Pioneering Affective-Semantic Bridge, explicitly validating the emergent capability of frozen VLMs to distill complex narrative intents into continuous emotion curves, thus bypassing the data bottleneck of traditional recognition. Second, we propose a data-efficient Dynamic Control mechanism via a token-level adapter. This module enables fine-grained tension modulation with minimal training overhead, achieving precise narrative alignment without disrupting the overarching musical structure. Finally, NarraScore achieves a synergistic balance between global style and local emotion, paving the way for fully automated, high-quality cinematic soundtrack generation.

\section{Related Work}

\subsection{Video to Music Generation}

The trajectory of video soundtrack generation has evolved from rule-based symbolic mappings to data-driven deep generative modeling. Early paradigms, represented by methods like CMT~\cite{di2021video} and Video2Music~\cite{kang2024video2music}, pioneered the Video-to-Music task by analyzing visual motion features to predict symbolic MIDI events. While foundational, these approaches often necessitated explicit user guidance to bridge the modality gap, resulting in limited expressive diversity and a heavy reliance on manual intervention.

The advent of audio language models marked a shift toward token-based generation. Recent frameworks such as MuVi~\cite{li2024muvi}, VMAS~\cite{lin2025vmas}, GVMGen~\cite{zuo2025gvmgen}, and VeM~\cite{tong2025video} leverage adapters to project dense video frames into latent conditions compatible with pre-trained backbones like MusicGen~\cite{copet2023simple}. Although effective for short clips, these methods predominantly rely on dense frame-level attention mechanisms. This design introduces severe scalability bottlenecks: for minute-level videos, the computational cost becomes prohibitive, and the dilution of attention leads to style drift and a loss of long-term coherence~\cite{tian2025vidmuse}.

To address these constraints, recent works have explored specific mechanisms for long-form generation. VidMuse~\cite{tian2025vidmuse} proposes specialized adapters to effectively model both long-term and short-term temporal features, employing a sliding-window inference strategy to ensure computational viability. JenBridge~\cite{yujenbridge} adopts a divide-and-conquer approach, segmenting videos for independent scoring and stitching them via transition techniques.

However, while these methods achieve acoustic continuity, they overlook the fundamental semantic shift in long-form content. Unlike short clips where a static mood suffices, long narratives possess evolving logic—tension rises, resolves, and shifts. By treating long videos merely as extended sequences, current solutions fail to capture these dynamic arcs, yielding monotonous background ambience rather than responsive, narrative-aligned soundtracks.Consequently, enabling the soundtrack to dynamically evolve in resonance with the narrative trajectory—transcending mere acoustic continuity—stands as the pivotal challenge for advancing video scoring toward professional-grade viability.

\subsection{Emotion-Driven Music Generation}

Given the intrinsic link between visual storytelling and musical expression, leveraging emotion as a conditioning signal has become a focal point. Early approaches primarily relied on discrete classification or global mapping. Video2Music~\cite{kang2024video2music} utilized CLIP for frame-level emotion classification, while EMSYNC~\cite{sulun2025video} advanced this by employing a psychology-driven mapping mechanism to translate discrete categorical predictions into continuous Valence and Arousal (VA) values. Despite these efforts, these methods rely heavily on surface-level visual semantic analysis, where the inherent accuracy limitations of standard CLIP-based classifiers often lead to coarse or noisy affective guidance.

To improve semantic fidelity of emotional control, subsequent works have integrated more robust priors.Methods like M2UGen~\cite{liu2023m}, FilmComposer~\cite{xie2025filmcomposer}, and JenBridge~\cite{yujenbridge} leverage Large Language Models (LLMs) to analyze visual content and generate descriptive emotion captions. While these methods demonstrate proficiency in synthesizing accurate global emotion labels, they largely circumvent the use of continuous emotion curves. This avoidance stems primarily from the scarcity of continuous affective data and the prohibitive cost of fine-grained annotation. Yet, for long-form video soundtrack generation, such dense temporal control is indispensable for maintaining plot consistency.

Although approaches like VeM~\cite{tong2025video} and MTCV2M~\cite{wu2025controllable} attempt to incorporate fine-grained control, they are fundamentally limited by their reliance on extrinsic guidance. This requirement for manual intervention renders such paradigms unscalable for autonomous, large-scale applications. Consequently, the automated and parameter-efficient extraction of deep, continuous affective cues solely from visual narratives remains a critical gap in achieving robust, high-fidelity control.

\subsection{Emotion Recognition from Video}
Affective video analysis serves as the perceptual foundation for emotion-driven content generation. However, the high annotative burden of continuous emotion labeling has led to a persistent scarcity of large-scale, high-quality datasets. To mitigate this supervision bottleneck, recent paradigms have shifted towards exploiting the generalized semantic representations of Vision-Language Pre-training (VLP) models. Pioneering works have demonstrated that
\begin{figure*}[t]
    \centering
    \includegraphics[width=\linewidth]{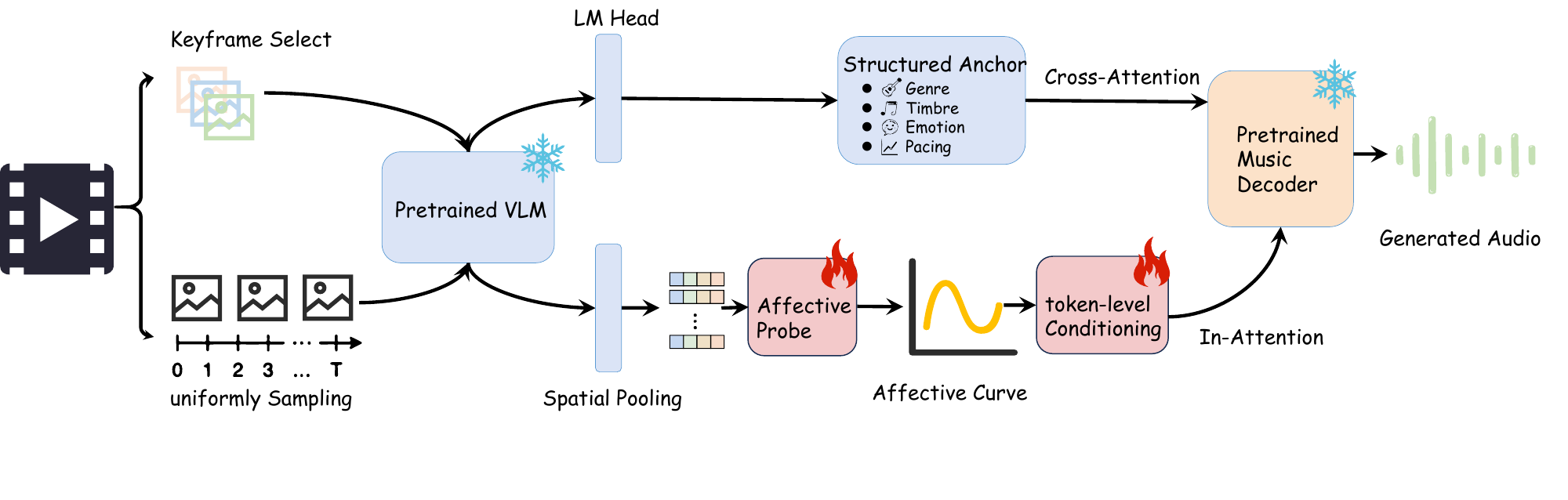}
    \caption{Overview of our framework}
    \label{fig:viz}
\end{figure*}
contrastive models, such as CLIP, possess robust zero-shot capabilities for static image emotion classification. Extending this to the temporal domain, methods like EmoCLIP~\cite{foteinopoulou2024emoclip, zhang2023learning} leverage CLIP embeddings to predict holistic video-level labels, while hybrid architectures integrate frozen visual features with learnable temporal modules to perform continuous Valence-Arousal (VA) regression.

Despite these advancements, directly deploying current State-of-the-Art (SOTA) affective recognition models as conditioning priors for music generation remains problematic. A fundamental domain misalignment exists within mainstream benchmarks like AFEW-VA~\cite{kossaifi2017afew}, Aff-Wild2~\cite{kollias2018aff}, and VEATIC~\cite{ren2024veatic}, which are predominantly face-centric, focusing on decoding the expressed emotion of actors via facial dynamics. In contrast, video soundtrack generation necessitates interpreting the induced affect—the overarching atmosphere and narrative tension perceived by viewers—rather than local facial cues. While datasets like LIRIS-ACCEDE~\cite{baveye2015liris} attempt to align with viewer perception, the limited scale across these benchmarks precludes the training of robust, data-hungry models. Furthermore, existing SOTA methods often lack the high-level semantic reasoning required to capture nuanced narrative shifts, resulting in noisy and temporally inconsistent predictions in complex, non-facial scenes. Consequently, relying on off-the-shelf emotion estimators as ground truth risks introducing significant error propagation, underscoring the imperative for approaches that leverage the deep reasoning capabilities of Large Language Models to infer contextually accurate emotion curves from limited supervision.

\section{Methodology}

\subsection{Problem Definition}
The primary objective of this work is to generate a musical sequence that mirrors the narrative progression of a long-form video. Formally, let $\mathcal{V} = \{v_1, \dots, v_{T_v}\}$ denote the input video sequence consisting of $T_v$ frames. The target output is a discrete acoustic sequence $\mathcal{A} \in \{1, \dots, N\}^{T_a \times K}$, derived from a neural audio codec~\cite{defossez2022high} with $K$ residual codebooks and a vocabulary size of $N$. Here, $T_a$ denotes the sequence length. The acoustic sequence is inherently dense, whereas the visual sequence $T_v$ is kept sparse align with the memory capacity limits when processing minute-level inputs.
Unlike short-clip generation, this task imposes dual constraints: global coherence for unified musical style, and local alignment for frame-level narrative synchrony.

\subsection{Overview}
To effectively reconcile these requirements, we introduce \textbf{NarraScore}, a hierarchical framework that disentangles the video-to-music generation task into two orthogonal dimensions: macro-scale atmospheric modeling and micro-scale tension tracking. Specifically, we bridge the semantic gap between the visual stream $\mathcal{V}$ and the acoustic domain $\mathcal{A}$ by introducing two decoupled priors: a global semantic anchor $\mathcal{S}_{global}$ and a frame-level affective trajectory $\mathcal{E}_{local}$. Consequently, the generative process is formulated as an auto regressive sequence prediction conditioned on these hierarchical cues:
\begin{equation}
p(\mathcal{A} \mid \mathcal{V}) = \prod_{t=1}^{T_a} p(a_t \mid a_{<t}, \mathcal{S}_{global}, \mathcal{E}_{local})
\end{equation}
As illustrated in Figure , the proposed architecture operationalizes this formulation through a cascaded Perception-Synthesis Pipeline. The workflow is anchored by a Unified Visual-Narrative Backbone, instantiated as a hybrid encoder integrating a Vision Transformer (ViT)~\cite{dosovitskiy2021image} front-end with a deep contextual reasoning stack. Leveraging a bifurcated projection mechanism, this unified system decomposes the raw visual stream into hierarchical control priors: it simultaneously regresses the frame-level affective trajectory $\mathcal{E}_{local}$ to delineate the micro-evolution of narrative tension, while abstracting the global semantic anchor $\mathcal{S}_{global}$ via a language modeling head to encapsulate the macro-atmospheric context. In the subsequent generative phase, these disentangled representations are integrated into a conditional acoustic transformer via distinct conditioning pathways, where the global anchor establishes the timbral and stylistic foundation, while the affective trajectory modulates the evolving narrative tension and musical dynamics.

\subsection{Narrative-Aware Affective Reasoning}
Deriving accurate emotion from video necessitates transcending static visual perception to capture evolving narrative dynamics. Rather than training a task-specific temporal encoder from scratch, we introduce a paradigm of Latent Semantic Probing. This approach functions as a unified spatiotemporal engine and orchestrates the rich reasoning priors of a frozen Vision-Language Model~\cite{zhang2025videollama} to lift raw pixel streams into a continuous affective manifold.
\paragraph{\textbf{Semantically-Anchored Temporal Alignment}}
To facilitate robust causal reasoning over long-form sequences, our framework adopts a strategy of Semantically-Anchored Temporal Alignment. This formulation leverages the architectural priors of the pre-trained backbone by employing linguistically grounded progression instead of rigid temporal embeddings. We discretize the video stream into a uniform 1Hz sequence to balance narrative granularity against the computational constraints of the backbone. These snapshots are interleaved with discrete semantic clocks $\tau_t$, formatted textually as ``Time: $t$s'', to strictly conform to the native interleaved schema of the model. The visual tokens $V_t$ retain their spatial fidelity via the intrinsic positional encoding mechanism. This effectively reconstructs the video as a linear causal sequence and establishes a continuous spatiotemporal context for subsequent affective reasoning.
\paragraph{\textbf{Instruction-Driven Semantic Steering}}
We bridge the domain gap between generalist object recognition and nuanced affective analysis by repurposing the instruction-following interface of the backbone. Rather than introducing external control modules, we optimize a system instruction $\mathcal{T}_{inst}$ to serve as a Semantic Primer. By prepending this primer to the aligned video sequence, we formulate the input representation $X$ as:
\begin{equation}
X = [\mathcal{T}_{inst}, \tau_1, V_1, \tau_2, V_2, \dots, \tau_T, V_T]
\end{equation}
Functionally, $\mathcal{T}{inst}$ modulates the self-attention mechanism to explicitly suppress the activation of low-level object enumeration patterns while activating high-level narrative reasoning pathways. This strategy steers the pre-trained capabilities towards analyzing narrative tension and emotional evolution to ensure that the extracted representations align with the affective task.
\paragraph{\textbf{Latent Affective Probing}}
To quantify the narrative tension distilled by the backbone, we introduce a lightweight probing head designed to extract the continuous affective trajectory $\mathcal{E}_{local}$. Let $Z_t \subset H^{(L)}$ denote the set of contextualized hidden states corresponding to the visual tokens of the $t$-th frame. We first aggregate these tokens via Spatial Average Pooling to obtain a holistic frame-level representation. This vector is then projected onto the Valence-Arousal plane~\cite{russell1980circumplex} via a Multi-Layer Perceptron:

\begin{equation}
e_t = \operatorname{Clip}_{[-1, 1]}\left( \operatorname{MLP} \left( \frac{1}{M} \sum_{z \in Z_t} z \right) \right)
\end{equation}
We freeze the massive backbone and train only this lightweight probe. This effectively distills the inherent ability of the VLM to correlate visual changes with explicit temporal progression into the specific task of affective regression and achieves high-fidelity prediction with minimal computational overhead.
\paragraph{\textbf{Optimization Objective}}
We employ a hybrid objective function combining L2 and L1 norms to calibrate the probing head against the target affective curves. This formulation balances the convergence stability provided by the Mean Squared Error with the robustness against outliers offered by the Mean Absolute Error. Let $\hat{e}_t \in \mathbb{R}^2$ denote the ground-truth Valence-Arousal vector for the $t$-th frame. The optimization objective $\mathcal{L}_{emo}$ is defined as:
\begin{equation}
\mathcal{L}{emo} = \frac{1}{T} \sum{t=1}^{T} \left( | e_t - \hat{e}_t |_2^2 + \lambda | e_t - \hat{e}_t |_1 \right)
\end{equation}
where $\lambda$ serves as a balancing coefficient.

\subsection{Holistic Musical Conceptualization}
Complementing the micro-level tension tracking, this component focuses on distilling the visual narrative into a global semantic anchor. The primary objective is to orchestrate the macro-level auditory identity that governs the overarching musical style and structural coherence. 

To bridge the semantic inference gap between visual features and musical concepts, we re-frame the interpretation task as a cross-modal sensory translation. Instead of relying on raw visual features, we leverage the reasoning priors of the VLM to identify the visual scene while exclusively describing the implied auditory imagery. This strategy achieves modality decoupling by explicitly suppressing references to specific visual objects and cinematography. It forces the model to abstract away from scene-specific visual nouns and project the content directly into acoustic descriptors semantically aligned with the downstream audio synthesis model~\cite{copet2023simple}.To guarantee robustness and consistency, we employ a structured instruction paradigm rather than free-form generation. We impose a strict schema that compels the VLM to synthesize a unified natural language description encompassing four essential musical dimensions: genre and stylistic context, instrumentation and timbral texture, emotional atmosphere, and rhythmic pacing. By projecting the visual content onto this predefined semantic subspace, we significantly reduce output variance and mitigate the ambiguity often associated with open-ended captioning. This structured constraint ensures that the generated control signals remain musically coherent and aligned with the intended aesthetic direction.
\begin{figure*}[t]
    \centering
    \includegraphics[width=\linewidth]{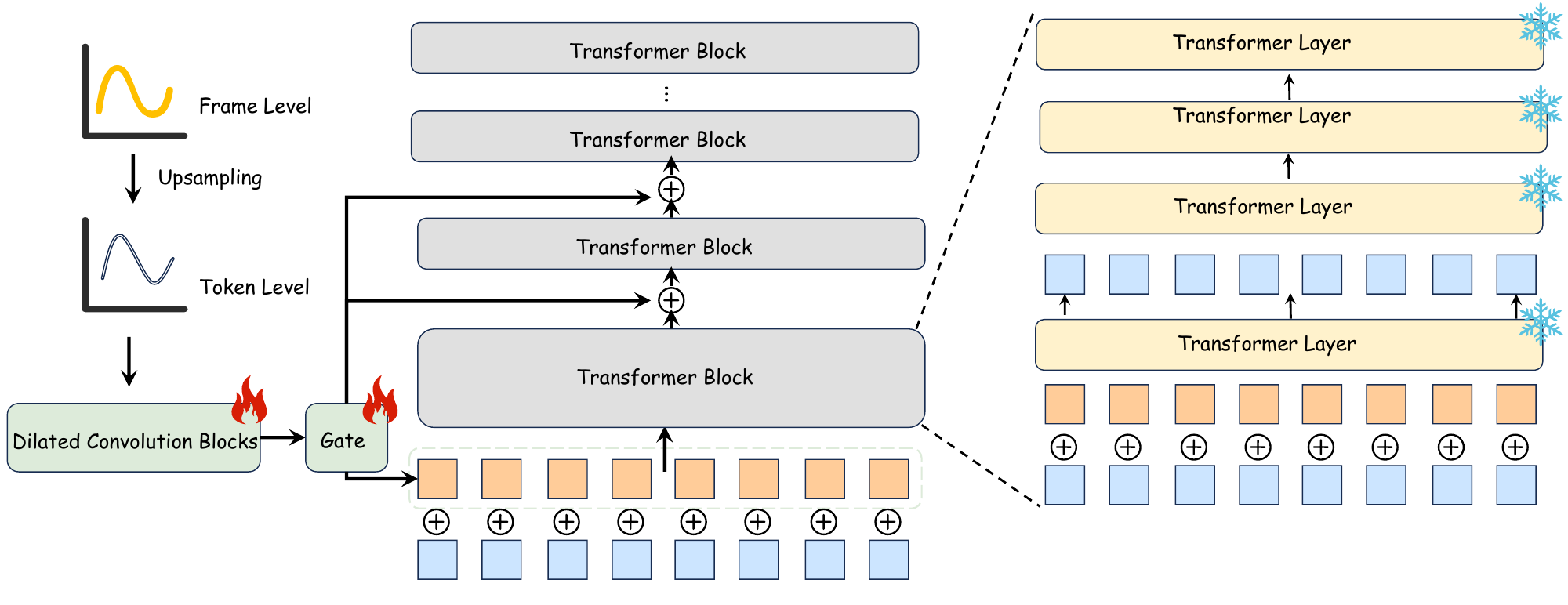}
    \caption{Our Method of Token-Wise Control Injection. }
    \label{fig:viz}
\end{figure*}
\subsection{Hierarchical Acoustic Synthesis}

The synthesis framework transforms the extracted narrative priors into a coherent acoustic waveform. To achieve this translation without compromising the high-fidelity generative distribution of the pre-trained music backbone~\cite{copet2023simple}, we employ a Dual-Stream Injection strategy. This mechanism aligns distinct control signals, comprising global style and local tension, with the hierarchical levels of the acoustic decoder.

\paragraph{\textbf{Explicit Semantic Bridging}}
The global semantic anchor $\mathcal{S}_{global}$ is integrated directly into the acoustic decoder via the pre-trained cross-attention mechanism. By utilizing the derived semantic description as the conditioning context, we steer the generative trajectory toward the target genre and atmosphere identified in the visual analysis phase. This standard conditioning approach ensures that the fundamental musical structure and instrumentation align with the narrative intent while preserving the original synthesis capabilities of the backbone without requiring invasive architectural modifications.

\paragraph{\textbf{Dense Affective Projection}}
A structural challenge in narrative alignment lies in the significant resolution discrepancy between the sparse visual emotion cues and the dense acoustic tokens. Directly injecting these sparse signals can lead to stepped and unnatural transitions in the musical output. To resolve this, we introduce a Temporal Super-resolution Adapter $\mathcal{F}_{\phi}$. This module first performs linear interpolation to align the discrete affective trajectory $\mathcal{E}_{local}$ with the target acoustic sequence length. Subsequently, a stack of dilated temporal convolutions\cite{bai2018empirical} is applied to smooth local jitter and expand the receptive field. This ensures that instantaneous emotional shifts are translated into fluid musical evolution. Formally, the dense control signal $C_{local} \in \mathbb{R}^{T_a \times D}$ is derived as:
\begin{equation}
C_{local} = \mathcal{F}_{\phi}(\operatorname{Interp}(\mathcal{E}_{local}))
\end{equation}
where $\mathcal{F}_{\phi}$ acts as a learnable mapping function that projects the 2-dimensional Valence-Arousal manifold into the $D$-dimensional latent space of the acoustic decoder.

\paragraph{\textbf{Token-Level Control Injection}}
To incorporate the dense affective features into the pre-trained backbone, we employ a residual modulation strategy~\cite{zhang2023adding}. Informed by observations in prior studies regarding the hierarchical distribution of information in music generation models~\cite{lan2024musicongen}, we restrict the injection of the control signal $C_{local}$ exclusively to the shallow Transformer blocks. This design choice implies that early layers are utilized to align the generation trajectory with the continuous affective constraints, thereby allowing the deeper layers to focus on the optimization of acoustic fidelity and harmonic coherence without interference. Formally, we apply a learnable additive bias to the hidden states $h^{(l)}_t$ at time step $t$ and layer $l$:
\begin{equation}
h^{(l)'}_t = h^{(l)}_t + \gamma \cdot C_{local, t}, \quad \forall l \in \{1, \dots, L_{shallow}\}
\end{equation}
Here, $\gamma$ is a learnable scalar initialized to zero to ensure the model retains its original distribution at the start of training. This selective injection scheme balances the trade-off between control precision and audio quality.

\paragraph{\textbf{Optimization Objective}}
To train the adapter module, we adhere to the native learning objective of the acoustic backbone. Keeping all parameters of the pre-trained autoregressive decoder frozen, we strictly optimize the adapter parameters $\phi$ and the gating scalars $\gamma$ using the standard autoregressive modeling objective. The process minimizes the Cross-Entropy loss between the predicted probability distribution and the ground-truth acoustic tokens, formulated as:
\begin{equation}
\mathcal{L}_{gen} = - \frac{1}{T_a} \sum_{t=1}^{T_a} \log p(a_t \mid a_{<t}, \mathcal{S}_{global}, C_{local})
\end{equation}
During the inference stage, the model utilizes the learned adapter to guide the backbone in generating the sequence of discrete acoustic tokens, which are subsequently reconstructed into the continuous high-fidelity waveform via the EnCodec~\cite{defossez2022high} decoder.

\subsection{Scalable Long-Form Inference}

To address the memory constraints inherent in processing minute-level sequences, we employ an overlapping sliding-window strategy. This approach enables the synthesis of temporally consistent soundtracks for long-form videos by combining global narrative abstraction with local context-dependent continuation.

\paragraph{\textbf{Global Semantic Reasoning}}
Prior to sequential processing, we derive the global semantic anchor $\mathcal{S}_{global}$ by leveraging the keyframe extraction capability of the VLM. This mechanism compresses the long-form visual content into a sparse sequence of representative frames and allows the model to perform holistic reasoning over the entire narrative arc within a limited context window. By distilling the global context into a unified stylistic description, we ensure that the subsequent music generation maintains a consistent thematic identity throughout the video duration.

\paragraph{\textbf{Continuous Affective Reasoning}}
In contrast to the global analysis, the extraction of local affective cues necessitates high temporal density. Consequently, we explicitly eschew temporal compression strategies to prevent the loss of fine-grained narrative dynamics and employ instead an overlapping sliding-window approach to extract the frame-level affective trajectory. The input video is processed in sequential windows that share a defined intersection. By utilizing this temporal overlap, we ensure that the prediction for the current window is contextually aligned with the preceding frames. This continuation approach guarantees that the extracted Valence and Arousal values evolve smoothly across window boundaries and effectively yields a continuous emotional curve despite the segmented processing.

\paragraph{\textbf{Autoregressive Acoustic Continuation}}
Similarly, the acoustic synthesis stage operates within this sliding-window framework. We maintain the global semantic anchor $\mathcal{S}_{global}$ as a stationary condition to preserve the overarching stylistic foundation while the dynamic narrative progression is governed by the continuous flow of the adapter signals. To ensure musical coherence at the overlap regions, we utilize the final sequence of acoustic tokens generated in the preceding window as the prompt prefix for the current window. This technique prompts the model to logically extend the musical phrase from the previous context. It guarantees seamless rhythmic continuity and acoustic causality throughout the long-form video.

\section{Experiments}
\label{sec:experiments}

\subsection{Experimental Setup}
\label{sec:setup}

\paragraph{\textbf{Datasets}}
We utilize publicly available benchmark datasets designed for continuous affective content analysis to train the video-to-emotion and emotion-to-music modules respectively.

For the video-to-emotion prediction task, we employ a continuous movie dataset designed for induced emotion prediction, which provides frame-level Valence-Arousal annotations capturing the emotional responses elicited in viewers during film watching. Unlike facial expression datasets that capture actors' portrayed emotions, this dataset annotates the holistic atmosphere and narrative tension perceived by audiences—precisely the type of induced affect required for background music generation. This alignment with the viewer's internal state is essential for generating background music that matches the narrative atmosphere.

For the emotion-to-music generation task, we utilize a music emotion dataset annotated for dynamic affective content, comprising excerpts and full songs with dense per-second valence-arousal labels. This dataset enables learning the mapping from emotional trajectories to musical characteristics, where continuous shifts in valence and arousal correspond to changes in harmony, tempo, and instrumentation. To ensure the model learns to control emotion independently of specific instrumentation, we aggregate metadata including genre and tags into textual captions. These captions are used as conditioning prompts during fine-tuning to decouple the representation of emotion from genre-specific stylistic conventions.

We apply a unified preprocessing pipeline to both datasets. The source separation model Demucs~\cite{defossez2019music} is utilized to suppress vocal tracks and isolate the instrumental background. We segment the continuous media streams into 30-second clips with a 15-second overlap to preserve temporal context. For the music generation task, we further refine the training set by discarding samples where the silence ratio exceeds 40\%. Following these procedures, the effective dataset sizes are approximately 884 minutes for the video dataset and 1351 minutes for the music dataset.

\paragraph{\textbf{Ethical Considerations}}
This research is conducted solely for academic and scientific purposes. The proposed method and experimental results are intended to advance the understanding of video-music alignment and affective computing. This work is not integrated into any commercial products or production systems.

\paragraph{\textbf{Implementation Details}}
Our framework integrates the pre-trained VideoLlama-3~\cite{zhang2025videollama} visual backbone and the MusicGen-Small~\cite{copet2023simple} acoustic decoder. The architecture incorporates two trainable modules to facilitate feature alignment, namely the Projector
% \begin{table}[t]
%     \centering
%     \small
%     \begin{tabular}{lcccc}
%         \toprule
%         \multirow{2}{*}{\textbf{Method}} & \multicolumn{2}{c}{\textbf{Valence}} & \multicolumn{2}{c}{\textbf{Arousal}} \\
%          & MSE & PCC & MSE & PCC \\
%         \midrule

%         \multicolumn{5}{l}{\textbf{Methods using multimodal inputs}} \\

%         GLA~\cite{sun2019gla} & 0.084 & 0.278 & 0.133 & 0.351 \\
%         Affect2MM~\cite{mittal2021affect} & 0.068 & - & 0.128 & - \\
%         RMN~\cite{zhang2022enlarging} & 0.068 & 0.471 & 0.124 & 0.468 \\
%         ET~\cite{park2022towards} & 0.092 & 0.358 & 0.134 & 0.308 \\
%         PGL-SUM~\cite{apostolidis2021combining} & 0.090 & 0.429 & 0.136 & 0.441 \\
%         RMN~\cite{zhang2022enlarging} & 0.073 & 0.515 & 0.124 & 0.441 \\
%         ULDA~\cite{liang2025learning} & 0.069 & 0.531 & 0.114 & 0.492 \\

%         \midrule
%         \multicolumn{5}{l}{\textbf{Methods using vision modality only}} \\

%         VideoCLIP~\cite{xu2021videoclip} & 0.142 & - & 0.151 & - \\
%         X-CLIP~\cite{ni2022expanding} & 0.133 & -  & 0.246 & - \\
%         EmotionCLIP~\cite{zhang2023learning} & 0.096 & - & 0.155 & - \\
%         ours & 0.078 & \textbf{0.543} & \textbf{0.091} & \textbf{0.527} \\

%         \bottomrule
%     \end{tabular}
%     \caption{Evaluation of continuous affective prediction on the video benchmark dataset. Metrics include MSE ($\downarrow$) and PCC ($\uparrow$).}
%     \label{tab:emotion_acc}
% \end{table}
and the Temporal Adapter.The Projector serves as a semantic interface. It accepts the extracted visual features $F_v \in \mathbb{R}^{T \times D_v}$ where $D_v$ denotes the visual embedding dimension. A two-layer MLP with GELU activation maps $F_v$ to the acoustic feature space $\mathbb{R}^{T \times D_a}$. We apply a dropout rate of 0.1 during this projection to regularize the mapping process.

The Temporal Adapter models long-term affective dependencies. It utilizes a dilated convolution layer on the sequence dimension to expand the temporal receptive field. This operation maintains the channel dimension $D_a$ and is followed by LeakyReLU activation and a linear layer to produce the final condition embeddings $C \in \mathbb{R}^{T \times D_a}$.

We employ a two-stage training strategy to ensure stability. The Projector is first trained for 150 epochs to align the static semantic features from $D_v$ to $D_a$. Subsequently, the Adapter is fine-tuned for 50 epochs to capture temporal dynamics.

\label{sec:exp_emotion}
\paragraph{\textbf{Baselines}}
To ensure a comprehensive evaluation, we benchmark \textbf{NarraScore} against a diverse set of five representative approaches spanning different generative paradigms. We first include M2UGEN~\cite{liu2023m} and video2music~\cite{kang2024video2music} as foundational multimodal frameworks that serve as established benchmarks in the field. To assess performance against the current state-of-the-art, we compare our method with VidMuse and GVMGEN, both of which represent the latest advancements in synchronized video-music synthesis. Furthermore, we construct a strong two-stage pipeline baseline named Caption2Music to evaluate the efficacy of disjoint modality processing. This baseline explicitly utilizes VideoLlama3-2B to generate detailed visual captions, which subsequently prompt MusicGen for audio synthesis. Including this cascaded model allows us to rigorously validate whether our proposed hierarchical injection strategy outperforms a naive combination of state-of-the-art vision and audio models.

% \subsection{Validating Affective Reasoning}
% To rigorously evaluate our emotion prediction module, we benchmark against representative approaches, including RBN, MMLGAN, AFRN, and TAM, on the video benchmark dataset.Adopting Mean Squared Error and Pearson Correlation Coefficient as metrics, our method establishes a new state-of-the-art across both Valence and Arousal dimensions as shown in Table X. Specifically, the significant improvement in Valence prediction stems from the powerful semantic understanding of pre-trained large models, which accurately map multimodal inputs to the continuous affective spectrum. A more critical challenge lies in Arousal regression, where the highly concentrated label distribution frequently drives existing models to degenerate into trivial mean-value prediction. In contrast, leveraging the extensive semantic priors and robust reasoning capabilities of large models, our approach effectively discerns subtle emotional fluctuations even within such a narrow range. This capability allows us to escape local optima and provides precise, fine-grained emotional guidance for the subsequent generation stage.

\begin{table}[ht]
    \centering
    \begin{tabular}{lccccc}
        \toprule
        Method & FAD ($\downarrow$) & FD ($\downarrow$) & KLD ($\downarrow$) & IB($\uparrow$)\\
        \midrule
        GT & 0 & 0 & 0 & 0.241\\
        VidMuse~\cite{tian2025vidmuse} & 2.459 & \textbf{29.946} & 0.734 & 0.202\\
        Video2Music~\cite{kang2024video2music} & 14.954 & 115.008 & 1.269 & 0.100\\
        GVMGen~\cite{zuo2025gvmgen} & \underline{2.362} & 41.466 & \underline{0.350} & \underline{0.213}\\
        M2UGEN~\cite{liu2023m} & 9.647 & 74.625 & 0.953 & 0.182\\
        \midrule
        NarraScore & \textbf{1.923} & \underline{36.411} & \textbf{0.320} & \textbf{0.219}\\
        \bottomrule
    \end{tabular}
    \caption{Quantitative comparison with state-of-the-art methods on objective metrics. $\uparrow$ indicates higher is better, $\downarrow$ indicates lower is better.\textbf{Bold} indicates the best performance, and \underline{underlined} indicates the second best.}
    \label{tab:main_results}
\end{table}

\subsection{Objective Evaluation}
\label{sec:exp_objective}

To quantitatively verify the efficacy of the proposed model, we conduct a comprehensive evaluation focusing on generation quality, fidelity, and diversity. Specifically, we employ Fréchet Audio Distance (FAD)~\cite{kilgour2019frechet}, Fréchet Distance (FD)~\cite{heusel2017gans}, and Kullback-Leibler Divergence (KLD) to measure the distributional discrepancies between the generated audio and the ground truth. To assess the variation in the synthesized samples, we utilize Density and Coverage metrics. Regarding cross-modal semantic consistency, we align with established baselines by incorporating the ImageBind score. However, it is pertinent to note that while ImageBind~\cite{girdhar2023imagebind} provides a metric for static semantic alignment, it may not fully capture the temporal progression and the intricate narrative flow inherent in video content. The quantitative comparisons are detailed in Table~\ref{tab:main_results}.

\subsection{Subjective Evaluation}
\label{sec:exp_subjective}
Recognizing human perception as the definitive benchmark for artistic generation, we conducted a user study involving 10 participants to evaluate the generated soundtracks. The assessment focused on five key dimensions: Emotional Dynamic Consistency (EDC),Global Style Matching (GSM),Long-term Coherence (LTC),Music Quality (MQ),Overall Preference (OP).

As presented in Table~\ref{tab:subjective_eval}, our method demonstrates superior performance across all metrics, establishing a significant lead over baseline approaches, particularly in Emotional Consistency. This advantage substantiates the model's capability to capture and articulate evolving affective content within the video. Conversely, the pipeline-based baseline (Caption2Music) exhibits marked deficiencies in visual–audio correspondence. This result suggests that exclusive reliance on textual captions creates an information bottleneck, filtering out the critical temporal and dynamic cues essential for precise audiovisual alignment.

\begin{table}[ht]
    \centering
    \small
    \begin{tabular}{lccccc}
        \toprule
        \textbf{Method} 
        & \textbf{EDC} 
        & \textbf{GSM} 
        & \textbf{LTC} 
        & \textbf{MQ} 
        & \textbf{OP} \\
        \midrule
        VidMuse~\cite{tian2025vidmuse}         & 2.29 & 2.49 & 2.49 & 2.13 & 2.18 \\
        Video2Music~\cite{kang2024video2music}     & 1.48 & 1.56 & 2.41 & \underline{3.39} & 1.88 \\
        GVMGen~\cite{zuo2025gvmgen}          & 1.65 & 1.69 & 1.41 & 1.64 & 1.34 \\
        Caption2Music   & \underline{2.66} & \underline{2.77} & \underline{2.88} & 3.18 & \underline{2.82} \\
        Ours            & \textbf{2.86} & \textbf{3.02} & \textbf{3.15} & \textbf{3.41} & \textbf{3.06} \\
        \bottomrule
    \end{tabular}
    \caption{Subjective Evaluation of Video-to-Music Generation on Long-form Videos.\textbf{Bold} indicates the best performance, and \underline{underlined} indicates the second best.}
    \label{tab:subjective_eval}
\end{table}

\begin{table}[ht]
    \centering
    \small
    \begin{tabular}{lccccc}
        \toprule
        \textbf{Method} 
        & \textbf{EDC} 
        & \textbf{GSM} 
        & \textbf{LTC} 
        & \textbf{MQ} 
        & \textbf{OVR} \\
        \midrule
        VidMuse~\cite{tian2025vidmuse}       & \underline{3.09} & \underline{3.27} & 3.03 & 3.00 & \underline{3.02} \\
        Video2Music~\cite{kang2024video2music}             & 1.66 & 1.68 & 2.38 & \underline{3.50} & 1.84 \\
        GVMGen~\cite{zuo2025gvmgen}        & 2.68 & 2.75 & 2.80 & 2.87 & 2.64 \\
        Caption2Music           & 2.83 & 3.01 & \underline{3.04} & 3.21 & 2.96 \\
        Ours                    & \textbf{3.11} & \textbf{3.36} & \textbf{3.24} & \textbf{3.52} & \textbf{3.32} \\
        \bottomrule
    \end{tabular}
    \caption{Subjective Evaluation of Video-to-Music Generation on Short- and Mid-length Videos.\textbf{Bold} indicates the best performance, and \underline{underlined} indicates the second best.}
    \label{tab:subjective_short}
\end{table}

The comparative analysis detailed in Tables \ref{tab:subjective_eval} and \ref{tab:subjective_short} reveals that performance disparities are intrinsically linked to the temporal modeling capabilities and conditioning mechanisms of each method. While baselines remain competitive in short- and mid-length scenarios due to limited temporal horizons, user feedback indicates a significant degradation in output quality when scaling to long-form videos. In these extended contexts, baseline methods frequently fail to sustain a stable musical trajectory, manifesting as narrative drift, inconsistent motifs, or disjointed segmentation. In contrast, our approach achieves the highest overall preference in both settings, with a widening margin in the long-form evaluation. This validates the effectiveness of our design in handling extended temporal dependencies, preserving a consistent global musical identity while adaptively responding to local visual cues.

Further analysis elucidates the specific limitations inherent in competing paradigms. The pipeline baseline, Caption2Music, highlights the limitations of text-only control, where coarse captions may capture the general mood but lack the temporal granularity required for fine-grained emotional transitions. Similarly, MIDI-based baselines such as Video2Music expose a dichotomy between acoustic fidelity and narrative alignment; despite receiving high ratings for audio texture, these tracks are frequently characterized by users as generic and weakly coupled to the visual storyline. Consequently, while short-term settings naturally narrow the performance gap by 
reducing the burden of long-range consistency, the sustained top ranking of our method across all metrics underscores that structural and affective coherence remain the defining factors in successful narrative music generation.

\subsection{Ablation Study}
To investigate the contribution of individual components and assess the framework's adaptability across different reasoning backbones, we conduct a comprehensive ablation study, as summarized in Table~\ref{tab:ablation}.
\begin{figure*}[ht]
    \centering
    \includegraphics[width=\linewidth]{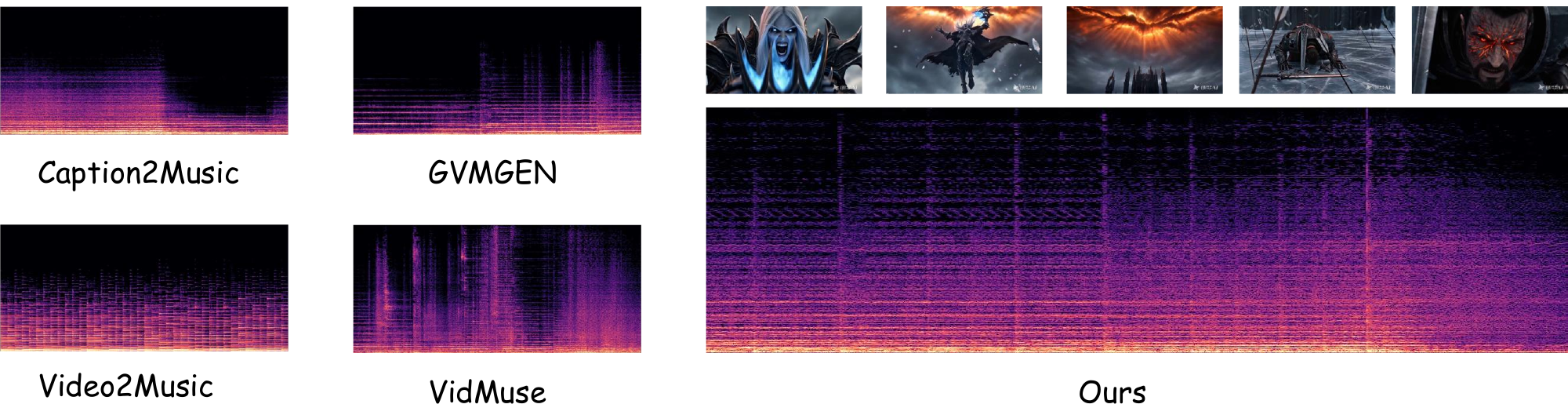}
    \caption{Visualization of the generated spectrograms and the corresponding narrative emotion curves.}
    \label{fig:mel_comparison}
\end{figure*}
The experimental results reveal distinct behaviors of the Holistic Musical Conceptualization (HMC) module across different scales of vision-language backbones. When employing lightweight VLMs as the captioner, the absence of HMC leads to noticeable performance degradation, as the model tends to produce literal descriptions of visual scenes without establishing meaningful connections to musical elements. However, an interesting phenomenon emerges when switching to powerful large-scale VLMs: the HMC module appears to constrain rather than enhance the model's capabilities. This suggests that advanced VLMs possess inherent cross-modal understanding abilities that allow them to directly infer appropriate musical characteristics from visual content without explicit high-level guidance. Regarding the adaptive in-attention mechanism, the results demonstrate that the 75\% injection ratio employed in NarraScore represents an optimal balance. Both reducing and increasing this proportion adversely affect performance, indicating that excessive narrative guidance may overwhelm the acoustic modeling capacity of the diffusion backbone while insufficient injection fails to provide adequate semantic alignment. Furthermore, the consistent improvements brought by the Narrative-Aware Affective Reasoning (NAR) mechanism across all backbone configurations validate its effectiveness as a model-agnostic component. Even when equipped with state-of-the-art VLMs, the incorporation of NAR continues to enhance music quality by bridging the affective gap between visual narratives and auditory expressions, underscoring the importance of explicit emotional reasoning in video-to-music generation tasks.

\subsection{Qualitative Analysis}
\label{sec:qualitative}
To qualitatively evaluate the acoustic characteristics and narrative alignment of the generated music, we perform a detailed comparative analysis of Mel-spectrograms produced by various models under identical video guidance. As illustrated in Figure~\ref{fig:mel_comparison}, the spectral architectures reveal fundamental differences in how each framework internalizes temporal dynamics and semantic logic.
Specifically, the Mel-spectrum of \textit{caption2music} is dominated by excessively smooth horizontal energy bands with a notable absence of vertical transients. This spectral stagnation indicates a lack of rhythmic pulses and dynamic fluctuations, which results in audio that fails to reflect the temporal energy shifts inherent in the video. In contrast, \textit{gvmgen} exhibits significant discontinuities in its spectral manifold characterized by fragmented energy distributions and abrupt temporal breaks. Such erratic spectral patterns suggest a failure in maintaining long-term acoustic coherence and a misalignment with the narrative progression of the visual input.
The results from \textit{v2m} further highlight these challenges because its spectrum appears overly monotonic and mechanical. The absence of structured harmonic variations and distinct beat markers implies a deficiency in narrative expression where the model produces a flat auditory output that remains unresponsive to visual climaxes. Furthermore, while \textit{vidmuse} generates content with high spectral density, it is plagued by a high concentration of stochastic noise in the high-frequency regions above 8000Hz. This is visualized as disordered pixel-like artifacts, which suggests that \textit{vidmuse} is confined to literal object-level associations instead of synthesizing a cohesive and narratively-driven soundtrack.

In contrast, our model demonstrates a superior spectral hierarchy that balances harmonic stability with rhythmic precision. The presence of clear fundamental frequencies alongside vertical onset markers indicates that our method successfully synchronizes discrete rhythmic events with the narrative arc. By capturing both the instantaneous motion cues and the global semantic flow, our model generates music that is not only acoustically clear but also contextually and emotionally resonant with the video content.

\begin{table}[th!]
    \centering
    \setlength{\tabcolsep}{6pt} % Adjust column spacing
    \begin{tabular}{lcccccc}
        \toprule
        Setting & FAD ($\downarrow$) & FD ($\downarrow$) & KLD ($\downarrow$) & IB ($\uparrow$)\\
        \midrule
        \textit{Component Analysis} & & & &  \\
        NarraScore & \textbf{1.923} & \underline{36.411} & \underline{0.320} & \textbf{0.219} \\
        \quad w/o HMC & 2.235 & \textbf{36.069} & 0.388 & \underline{0.203} \\
        \quad w/o NAR & 3.009 & 41.146 & 0.545 & 0.202\\
        \midrule
        \textit{In-attention Analysis} & & & & \\
        \quad 50\% Blocks & 2.021 & 37.282 & 0.353 & 0.196 \\
        \quad 100\% Blocks & \underline{1.964} & 36.503 & \textbf{0.318} & 0.200 \\
        \midrule
        \textit{Backbone Analysis} & & & & \\
        Gemini2.5pro~\cite{comanici2025gemini} & 2.322 & 33.752 & 0.376 & \textbf{0.226} \\
        Gemini (w/o HMC) & \textbf{1.906} & \textbf{31.299} & \textbf{0.320} & \underline{0.223}\\
        Gemini (w/o NAR) & 2.430 & 32.568 & 0.403 & 0.214\\
        Gemini (cap-only) & \underline{2.002} & \underline{32.239} & \underline{0.324} & 0.203\\
        \bottomrule
    \end{tabular}
    \caption{Ablation study on key components and different LLM backbones. \textbf{NAR} denotes Narrative-Aware Affective Reasoning.\textbf{Bold} indicates the best performance, and \underline{underlined} indicates the second best.}
    % \textbf{HMC} denotes Holistic Musical Conceptualization, and 
    \label{tab:ablation}
\end{table}

\section{Conclusion}
We present NarraScore, pioneering the first direct emotional control pathway from video narratives to musical dynamics. Our findings demonstrate that fine-tuning small mLLMs with limited labeled data enables robust continuous temporal regression. Simultaneously, we confirm that a lightweight adapter is sufficient to steer the complex acoustic backbone. By reconciling global style with local tension, NarraScore achieves state-of-the-art coherence, establishing a strong baseline for autonomous soundtrack generation.

\section{Limitations}
Current limitations stem from the limited temporal granularity of the affective control, which precludes frame-perfect synchronization with rapid visual events. Additionally, the cascaded design risks error propagation from upstream affective reasoning. Future work will focus on end-to-end joint optimization to mitigate these dependencies and explore knowledge distillation to reduce the computational latency of the visual backbone.
% \section{Acknowledgments}

% Identification of funding sources and other support, and thanks to
% individuals and groups that assisted in the research and the
% preparation of the work should be included in an acknowledgment
% section, which is placed just before the reference section in your
% document.

% \begin{acks}
% \end{acks}

\newpage

\bibliographystyle{ACM-Reference-Format}
\bibliography{sample-base}

@article{dasovich2022exploring,
  title={Exploring music video experiences and their influence on music perception},
  author={Dasovich-Wilson, Johanna N and Thompson, Marc and Saarikallio, Suvi},
  journal={Music \& Science},
  volume={5},
  pages={20592043221117651},
  year={2022},
  publisher={SAGE Publications Sage UK: London, England}
}

@article{ma2022research,
  title={Research on the effect of different types of short music videos on viewers' psychological emotions},
  author={Ma, Lin},
  journal={Frontiers in public health},
  volume={10},
  pages={992200},
  year={2022},
  publisher={Frontiers Media SA}
}

@article{millet2021soundtrack,
  title={Soundtrack design: The impact of music on visual attention and affective responses},
  author={Millet, Barbara and Chattah, Juan and Ahn, Soyeon},
  journal={Applied ergonomics},
  volume={93},
  pages={103301},
  year={2021},
  publisher={Elsevier}
}

@inproceedings{di2021video,
  title={Video background music generation with controllable music transformer},
  author={Di, Shangzhe and Jiang, Zeren and Liu, Si and Wang, Zhaokai and Zhu, Leyan and He, Zexin and Liu, Hongming and Yan, Shuicheng},
  booktitle={Proceedings of the 29th ACM International Conference on Multimedia},
  pages={2037--2045},
  year={2021}
}

@inproceedings{gan2020foley,
  title={Foley music: Learning to generate music from videos},
  author={Gan, Chuang and Huang, Deng and Chen, Peihao and Tenenbaum, Joshua B and Torralba, Antonio},
  booktitle={European Conference on Computer Vision},
  pages={758--775},
  year={2020},
  organization={Springer}
}

@inproceedings{su2024v2meow,
  title={V2meow: Meowing to the visual beat via video-to-music generation},
  author={Su, Kun and Li, Judith Yue and Huang, Qingqing and Kuzmin, Dima and Lee, Joonseok and Donahue, Chris and Sha, Fei and Jansen, Aren and Wang, Yu and Verzetti, Mauro and others},
  booktitle={Proceedings of the AAAI Conference on Artificial Intelligence},
  volume={38},
  number={5},
  pages={4952--4960},
  year={2024}
}

@article{kang2024video2music,
  title={Video2music: Suitable music generation from videos using an affective multimodal transformer model},
  author={Kang, Jaeyong and Poria, Soujanya and Herremans, Dorien},
  journal={Expert Systems with Applications},
  volume={249},
  pages={123640},
  year={2024},
  publisher={Elsevier}
}

@article{huang2018music,
  title={Music transformer},
  author={Huang, Cheng-Zhi Anna and Vaswani, Ashish and Uszkoreit, Jakob and Shazeer, Noam and Simon, Ian and Hawthorne, Curtis and Dai, Andrew M and Hoffman, Matthew D and Dinculescu, Monica and Eck, Douglas},
  journal={arXiv preprint arXiv:1809.04281},
  year={2018}
}

@article{agostinelli2023musiclm,
  title={Musiclm: Generating music from text},
  author={Agostinelli, Andrea and Denk, Timo I and Borsos, Zal{\'a}n and Engel, Jesse and Verzetti, Mauro and Caillon, Antoine and Huang, Qingqing and Jansen, Aren and Roberts, Adam and Tagliasacchi, Marco and others},
  journal={arXiv preprint arXiv:2301.11325},
  year={2023}
}

@article{copet2023simple,
  title={Simple and controllable music generation},
  author={Copet, Jade and Kreuk, Felix and Gat, Itai and Remez, Tal and Kant, David and Synnaeve, Gabriel and Adi, Yossi and D{\'e}fossez, Alexandre},
  journal={Advances in Neural Information Processing Systems},
  volume={36},
  pages={47704--47720},
  year={2023}
}

@inproceedings{zuo2025gvmgen,
  title={Gvmgen: A general video-to-music generation model with hierarchical attentions},
  author={Zuo, Heda and You, Weitao and Wu, Junxian and Ren, Shihong and Chen, Pei and Zhou, Mingxu and Lu, Yujia and Sun, Lingyun},
  booktitle={Proceedings of the AAAI Conference on Artificial Intelligence},
  volume={39},
  number={21},
  pages={23099--23107},
  year={2025}
}

@inproceedings{lin2025vmas,
  title={Vmas: Video-to-music generation via semantic alignment in web music videos},
  author={Lin, Yan-Bo and Tian, Yu and Yang, Linjie and Bertasius, Gedas and Wang, Heng},
  booktitle={2025 IEEE/CVF Winter Conference on Applications of Computer Vision (WACV)},
  pages={1155--1165},
  year={2025},
  organization={IEEE}
}

@article{tong2022videomae,
  title={Videomae: Masked autoencoders are data-efficient learners for self-supervised video pre-training},
  author={Tong, Zhan and Song, Yibing and Wang, Jue and Wang, Limin},
  journal={Advances in neural information processing systems},
  volume={35},
  pages={10078--10093},
  year={2022}
}

@inproceedings{wang2023videomae,
  title={Videomae v2: Scaling video masked autoencoders with dual masking},
  author={Wang, Limin and Huang, Bingkun and Zhao, Zhiyu and Tong, Zhan and He, Yinan and Wang, Yi and Wang, Yali and Qiao, Yu},
  booktitle={Proceedings of the IEEE/CVF conference on computer vision and pattern recognition},
  pages={14549--14560},
  year={2023}
}

@inproceedings{feichtenhofer2019slowfast,
  title={Slowfast networks for video recognition},
  author={Feichtenhofer, Christoph and Fan, Haoqi and Malik, Jitendra and He, Kaiming},
  booktitle={Proceedings of the IEEE/CVF international conference on computer vision},
  pages={6202--6211},
  year={2019}
}

@article{ji2025comprehensive,
  title={A Comprehensive Survey on Generative AI for Video-to-Music Generation},
  author={Ji, Shulei and Wu, Songruoyao and Wang, Zihao and Li, Shuyu and Zhang, Kejun},
  journal={arXiv preprint arXiv:2502.12489},
  year={2025}
}

@article{zhang2023video,
  title={Video-llama: An instruction-tuned audio-visual language model for video understanding},
  author={Zhang, Hang and Li, Xin and Bing, Lidong},
  journal={arXiv preprint arXiv:2306.02858},
  year={2023}
}

@article{cheng2024videollama,
  title={Videollama 2: Advancing spatial-temporal modeling and audio understanding in video-llms},
  author={Cheng, Zesen and Leng, Sicong and Zhang, Hang and Xin, Yifei and Li, Xin and Chen, Guanzheng and Zhu, Yongxin and Zhang, Wenqi and Luo, Ziyang and Zhao, Deli and others},
  journal={arXiv preprint arXiv:2406.07476},
  year={2024}
}

@article{zhang2025videollama,
  title={Videollama 3: Frontier multimodal foundation models for image and video understanding},
  author={Zhang, Boqiang and Li, Kehan and Cheng, Zesen and Hu, Zhiqiang and Yuan, Yuqian and Chen, Guanzheng and Leng, Sicong and Jiang, Yuming and Zhang, Hang and Li, Xin and others},
  journal={arXiv preprint arXiv:2501.13106},
  year={2025}
}

@misc{openai2024sora,
  title        = {Video generation models as world simulators},
  author       = {{OpenAI}},
  year         = {2024},
  howpublished = {\url{https://openai.com/index/video-generation-models-as-world-simulators/}}
}

@misc{deepmind2025veo3,
  title        = {Veo 3 Tech Report},
  author       = {{Google DeepMind}},
  year         = {2025},
  howpublished = {\url{https://storage.googleapis.com/deepmind-media/veo/Veo-3-Tech-Report.pdf}}
}

@article{wiedemer2025video,
  title={Video models are zero-shot learners and reasoners},
  author={Wiedemer, Thadd{\"a}us and Li, Yuxuan and Vicol, Paul and Gu, Shixiang Shane and Matarese, Nick and Swersky, Kevin and Kim, Been and Jaini, Priyank and Geirhos, Robert},
  journal={arXiv preprint arXiv:2509.20328},
  year={2025}
}

@article{liu2025improving,
  title={Improving video generation with human feedback},
  author={Liu, Jie and Liu, Gongye and Liang, Jiajun and Yuan, Ziyang and Liu, Xiaokun and Zheng, Mingwu and Wu, Xiele and Wang, Qiulin and Xia, Menghan and Wang, Xintao and others},
  journal={arXiv preprint arXiv:2501.13918},
  year={2025}
}

@inproceedings{zhang2023adding,
  title={Adding conditional control to text-to-image diffusion models},
  author={Zhang, Lvmin and Rao, Anyi and Agrawala, Maneesh},
  booktitle={Proceedings of the IEEE/CVF international conference on computer vision},
  pages={3836--3847},
  year={2023}
}

@article{lan2024musicongen,
  title={Musicongen: Rhythm and chord control for transformer-based text-to-music generation},
  author={Lan, Yun-Han and Hsiao, Wen-Yi and Cheng, Hao-Chung and Yang, Yi-Hsuan},
  journal={arXiv preprint arXiv:2407.15060},
  year={2024}
}

@article{rouditchenko2020avlnet,
  title={Avlnet: Learning audio-visual language representations from instructional videos},
  author={Rouditchenko, Andrew and Boggust, Angie and Harwath, David and Chen, Brian and Joshi, Dhiraj and Thomas, Samuel and Audhkhasi, Kartik and Kuehne, Hilde and Panda, Rameswar and Feris, Rogerio and others},
  journal={arXiv preprint arXiv:2006.09199},
  year={2020}
}

@article{li2024muvi,
  title={Muvi: Video-to-music generation with semantic alignment and rhythmic synchronization},
  author={Li, Ruiqi and Zheng, Siqi and Cheng, Xize and Zhang, Ziang and Ji, Shengpeng and Zhao, Zhou},
  journal={arXiv preprint arXiv:2410.12957},
  year={2024}
}

@article{tong2025video,
  title={Video Echoed in Music: Semantic, Temporal, and Rhythmic Alignment for Video-to-Music Generation},
  author={Tong, Xinyi and Zhu, Yiran and Chen, Jishang and Zhan, Chunru and Wang, Tianle and Zhang, Sirui and Liu, Nian and Ge, Tiezheng and Xu, Duo and Jin, Xin and others},
  journal={arXiv preprint arXiv:2511.09585},
  year={2025}
}

@inproceedings{tian2025vidmuse,
  title={Vidmuse: A simple video-to-music generation framework with long-short-term modeling},
  author={Tian, Zeyue and Liu, Zhaoyang and Yuan, Ruibin and Pan, Jiahao and Liu, Qifeng and Tan, Xu and Chen, Qifeng and Xue, Wei and Guo, Yike},
  booktitle={Proceedings of the Computer Vision and Pattern Recognition Conference},
  pages={18782--18793},
  year={2025}
}

@article{yujenbridge,
  title={JenBridge: Adaptive Long-Form Video Soundtracking across Scene Transition},
  author={Yu, Jiashuo and Yao, Yao and Chen, Boyu and Wang, Alex}
}

@article{sulun2025video,
  title={Video Soundtrack Generation by Aligning Emotions and Temporal Boundaries},
  author={Sulun, Serkan and Viana, Paula and Davies, Matthew EP},
  journal={arXiv preprint arXiv:2502.10154},
  year={2025}
}

@article{liu2023m,
  title={{M$^2$UGen}: Multi-modal Music Understanding and Generation with the Power of Large Language Models},
  author={Liu, Shansong and Hussain, Atin Sakkeer and Wu, Qilong and Sun, Chenshuo and Shan, Ying},
  journal={arXiv preprint arXiv:2311.11255},
  year={2023}
}

@inproceedings{xie2025filmcomposer,
  title={FilmComposer: LLM-Driven Music Production for Silent Film Clips},
  author={Xie, Zhifeng and He, Qile and Zhu, Youjia and He, Qiwei and Li, Mengtian},
  booktitle={Proceedings of the Computer Vision and Pattern Recognition Conference},
  pages={13519--13528},
  year={2025}
}

@inproceedings{wu2025controllable,
  title={Controllable video-to-music generation with multiple time-varying conditions},
  author={Wu, Junxian and You, Weitao and Zuo, Heda and Zhang, Dengming and Chen, Pei and Sun, Lingyun},
  booktitle={Proceedings of the 33rd ACM International Conference on Multimedia},
  pages={10427--10436},
  year={2025}
}

@inproceedings{foteinopoulou2024emoclip,
  title={Emoclip: A vision-language method for zero-shot video facial expression recognition},
  author={Foteinopoulou, Niki Maria and Patras, Ioannis},
  booktitle={2024 IEEE 18th International Conference on Automatic Face and Gesture Recognition (FG)},
  pages={1--10},
  year={2024},
  organization={IEEE}
}

@inproceedings{zhang2023learning,
  title={Learning emotion representations from verbal and nonverbal communication},
  author={Zhang, Sitao and Pan, Yimu and Wang, James Z},
  booktitle={Proceedings of the IEEE/CVF Conference on Computer Vision and Pattern Recognition},
  pages={18993--19004},
  year={2023}
}

@article{kossaifi2017afew,
  title={AFEW-VA database for valence and arousal estimation in-the-wild},
  author={Kossaifi, Jean and Tzimiropoulos, Georgios and Todorovic, Sinisa and Pantic, Maja},
  journal={Image and Vision Computing},
  volume={65},
  pages={23--36},
  year={2017},
  publisher={Elsevier}
}

@article{kollias2018aff,
  title={Aff-wild2: Extending the aff-wild database for affect recognition},
  author={Kollias, Dimitrios and Zafeiriou, Stefanos},
  journal={arXiv preprint arXiv:1811.07770},
  year={2018}
}

@inproceedings{ren2024veatic,
  title={Veatic: Video-based emotion and affect tracking in context dataset},
  author={Ren, Zhihang and Ortega, Jefferson and Wang, Yifan and Chen, Zhimin and Guo, Yunhui and Yu, Stella X and Whitney, David},
  booktitle={Proceedings of the IEEE/CVF Winter Conference on Applications of Computer Vision},
  pages={4467--4477},
  year={2024}
}

@article{baveye2015liris,
  title={LIRIS-ACCEDE: A video database for affective content analysis},
  author={Baveye, Yoann and Dellandrea, Emmanuel and Chamaret, Christel and Chen, Liming},
  journal={IEEE Transactions on Affective Computing},
  volume={6},
  number={1},
  pages={43--55},
  year={2015},
  publisher={IEEE}
}

@article{defossez2022high,
  title={High fidelity neural audio compression},
  author={D{\'e}fossez, Alexandre and Copet, Jade and Synnaeve, Gabriel and Adi, Yossi},
  journal={arXiv preprint arXiv:2210.13438},
  year={2022}
}

@misc{dosovitskiy2021image,
      title={An Image is Worth 16x16 Words: Transformers for Image Recognition at Scale}, 
      author={Alexey Dosovitskiy and Lucas Beyer and Alexander Kolesnikov and Dirk Weissenborn and Xiaohua Zhai and Thomas Unterthiner and Mostafa Dehghani and Matthias Minderer and Georg Heigold and Sylvain Gelly and Jakob Uszkoreit and Neil Houlsby},
      year={2021},
      eprint={2010.11929},
      archivePrefix={arXiv},
      primaryClass={cs.CV},
      url={https://arxiv.org/abs/2010.11929}, 
}

@article{russell1980circumplex,
  title={A circumplex model of affect.},
  author={Russell, James A},
  journal={Journal of personality and social psychology},
  volume={39},
  number={6},
  pages={1161},
  year={1980},
  publisher={American Psychological Association}
}

@misc{bai2018empirical,
      title={An Empirical Evaluation of Generic Convolutional and Recurrent Networks for Sequence Modeling}, 
      author={Shaojie Bai and J. Zico Kolter and Vladlen Koltun},
      year={2018},
      eprint={1803.01271},
      archivePrefix={arXiv},
      primaryClass={cs.LG},
      url={https://arxiv.org/abs/1803.01271}, 
}

@article{defossez2019music,
  title={Music source separation in the waveform domain},
  author={D{\'e}fossez, Alexandre and Usunier, Nicolas and Bottou, L{\'e}on and Bach, Francis},
  journal={arXiv preprint arXiv:1911.13254},
  year={2019}
}

@misc{kilgour2019frechet,
      title={Fr\'echet Audio Distance: A Metric for Evaluating Music Enhancement Algorithms}, 
      author={Kevin Kilgour and Mauricio Zuluaga and Dominik Roblek and Matthew Sharifi},
      year={2019},
      eprint={1812.08466},
      archivePrefix={arXiv},
      primaryClass={eess.AS},
      url={https://arxiv.org/abs/1812.08466}, 
}

@misc{girdhar2023imagebind,
      title={ImageBind: One Embedding Space To Bind Them All}, 
      author={Rohit Girdhar and Alaaeldin El-Nouby and Zhuang Liu and Mannat Singh and Kalyan Vasudev Alwala and Armand Joulin and Ishan Misra},
      year={2023},
      eprint={2305.05665},
      archivePrefix={arXiv},
      primaryClass={cs.CV},
      url={https://arxiv.org/abs/2305.05665}, 
}

@article{heusel2017gans,
  title={Gans trained by a two time-scale update rule converge to a local nash equilibrium},
  author={Heusel, Martin and Ramsauer, Hubert and Unterthiner, Thomas and Nessler, Bernhard and Hochreiter, Sepp},
  journal={Advances in neural information processing systems},
  volume={30},
  year={2017}
}

@article{comanici2025gemini,
  title={Gemini 2.5: Pushing the frontier with advanced reasoning, multimodality, long context, and next generation agentic capabilities},
  author={Comanici, Gheorghe and Bieber, Eric and Schaekermann, Mike and Pasupat, Ice and Sachdeva, Noveen and Dhillon, Inderjit and Blistein, Marcel and Ram, Ori and Zhang, Dan and Rosen, Evan and others},
  journal={arXiv preprint arXiv:2507.06261},
  year={2025}
}

% \input{chapters/Appendix}

\end{document}